\documentclass[american,aps,pra,reprint, superscriptaddress]{revtex4-1}
\usepackage{amsmath,amscd,amsthm,amssymb}
\usepackage{mathrsfs}
\usepackage{graphicx}
\usepackage{epsf,epstopdf}
\usepackage{caption3}
\usepackage[all]{xy}
\usepackage{tikz}

\usepackage[unicode=true,pdfusetitle, bookmarks=true,bookmarksnumbered=false,bookmarksopen=false, breaklinks=false,pdfborder={0 0 0},backref=false,colorlinks=false] {hyperref}
\hypersetup{ colorlinks,linkcolor=myurlcolor,citecolor=myurlcolor,urlcolor=myurlcolor}
\usepackage{graphics,epstopdf,graphicx, amsthm, amsmath, amssymb, times, braket, color, bm}
\usepackage[up]{subfigure}
\usepackage{cleveref}

\definecolor{myurlcolor}{rgb}{0,0,0.7}

\def\be{\begin{equation}}
\def\ee{\end{equation}}
\def\bea{\begin{eqnarray*}}
\def\eea{\end{eqnarray*}}
\def\ot{\otimes}

\theoremstyle{plain}
\newtheorem{thrm}{\protect\theoremname}

\newtheorem{defi}[thrm]{Definition}
\newtheorem{coro}[thrm]{Corollary}
\newtheorem{fact}[thrm]{Fact}

\newtheorem{lem}[thrm]{Lemma}

\providecommand{\theoremname}{Theorem}
\newcommand{\inner}[2]{\langle #1 , #2\rangle}
\newcommand{\iinner}[2]{\langle #1 | #2\rangle}
\newcommand{\out}[2]{| #1\rangle\langle #2 |}
\DeclareMathOperator{\trace}{tr}
\newcommand{\ptr}[2]{\trace_{#1}({#2})}
\newcommand{\tr}[1]{\ptr{}{#1}}
\newcommand{\id}{\mathbb{I}}

\newcommand*{\myproofname}{Proof}

\makeatother

\def\cC{\mathcal{C}}\def\cD{\mathcal{D}}\def\cE{\mathcal{E}}
\def\cF{\mathcal{F}}\def\cH{\mathcal{H}}

\def\cR{\mathcal{R}}

\def\rD{\mathrm{D}}
\def\rF{\mathrm{F}}

\def\rR{\mathrm{R}}

\theoremstyle{definition}

\theoremstyle{remark}

\graphicspath{{figs/}}

\begin{document}
 \author{Sunho Kim}
 \email{kimshanhao@126.com}
 \affiliation{School of Mathematical Sciences, Harbin Engineering University, Harbin 150001, People's Republic of China}

  \author{Chunhe Xiong}
 \email{xiongchunhe@csu.edu.cn}
 \affiliation{School of Mathematics and Statistics, Central South University, Changsha 410000, People's Republic of China}

 \author{Junde Wu}
 \email{wjd@zju.edu.cn: Corresponding author}
 \affiliation{School of Mathematical Sciences, Zhejiang University, Hangzhou 310027, People's Republic of China}

\title{Relative Quantum Resource Theory and Operational Applications in Subchannel Discrimination}
\begin{abstract}
A central problem in quantum resource theory is to give operational meaning to quantum resources that can provide clear advantages in certain physical tasks compared to the convex set of resource-free states. We propose to extend this basic principle by defining the relative superiority of resources over a specific convex set of resource states, also provide a relative advantage in physical tasks based on this extended principle. This allows the generalized robustness measure to quantify the relative maximal advantage due to a given resource state over a specific convex set of resource states in the subchannel discrimination, thereby showing that the operational interpretation of resource measures also holds in a relative perspective.
In addition, we offer a new framework for defining the deficiency of a given state in physical tasks compared to the set of maximum resource states. The geometric measure we provide satisfies the conditions of the framework for quantum coherence and entanglement, and it accurately quantifies the minimal disadvantage due to a given state compared to maximum resource states in the subchannel discrimination in certain situations. These two extensions and new interpretations expand the scope of quantum resource theories and provide a more comprehensive operational interpretation.

\end{abstract}
\maketitle

{\it Introduction.--} Quantum resource theory plays an important role in implementing quantum information and quantum computation tasks and provides a versatile and robust framework for studying various phenomena in quantum theory. From quantum entanglement to quantum coherence, resource theory is responsible for quantifying various effects in quantum domains \cite{Bennett, Plenio, Regula}, developing new detection protocols \cite{Horodecki1, Schaetz, Hou}, and identifying processes that optimize usage for a given application \cite{Roa}. Quantum resource theory has become a powerful and reliable tool.

Various theories and experiments of quantum resources are progressing in recent decades, and many reviews have been made such as entanglement \cite{Horodecki2}, coherence \cite{Streltsov}, quantum reference frames and asymmetry \cite{Bartlett}, quantum thermodynamics \cite{Gour}, nonlocality \cite{Luo, Brunner}, nongaussianity \cite{Weedbrook}, and quantum correlations \cite{Luo1, Modi}. In particular, for some resources, such as quantum entanglement and correlations, several studies have been conducted to characterize and quantify for the quantum states of multipartite systems beyond bipartites \cite{Mintert, Vicente, Hou1}.
In addition, the dynamical resource theory is also being systematized \cite{Gour1, Gour2, Saxena}. 

In resource theory, the set of free states is an essential component. It represents a states set are "easy to prepare" or "provided for free", their properties are governed by classical physics. The states outside of this set are called resource states. A general and intuitive assumption is that the set of free states should be convex and closed. These properties reflect the natural properties of many physical environments. In principle, free states are established for most quantum resources, and the quantifiers of those resources are defined based on the set of all free states \cite{Regula, Chitambar}. This gives the following framework for quantifying quantum resources. Let $\cH$ be a finite dimensional Hilbert space with $d = \dim{\cH}$. In general, in resource theory for states, measures for quantum resources are defined as satisfying the following conditions:

(R1) faithful: $\rR(\rho)\geq 0$, and $\rR(\rho)=0$ if and only if $\rho \in \cF(\cH)$ where $\cF(\cH)$ is the set of all free states;

(R2a) monotonicity under any free operation $\Sigma$: $\rR(\Sigma(\rho)) \leq \rR(\rho)$ for any free operations $\Sigma$,

or (R2b) monotonicity under selective measurement $\{K_n\}$ : $\sum_np_n\rR(\sigma_n)\leq \rR(\rho)$, where $\sigma_n = K_n\rho K_n^\dagger/p_n$ with $p_n = \tr{K_n\rho K_n^\dagger}$;

(R3) convexity: $\rR(\sum_iq_i\rho_i) \leq \sum_iq_i\rR(\rho_i)$. \\
Note that  (R2b, R3) imply (R2a). When (R1, R2b) is satisfied, it is called a monotone, while when (R1-R3) is satisfied, it is called a measure (In this paper, we will not discuss the weak measures that satisfy only R1, R2a and R3).
To date, various quantifier protocols have been presented, including: robustness of resource \cite{Vidal, Harrow, Napoli}, distance-based resource\cite{Vedral, Baumgratz}, relative entropy of resource \cite{Renyi, Eisert, Winter}, resource distillation \cite{Bennett, Winter, Bravyi}, and weight of resource \cite{Elitzur, Lewenstein, Skrzypczyk}.

What we emphasize in this letter is the fact that certain quantifiers can characterize physical tasks that provide an explicit advantage over all resource-free states by providing operational meaning to a given resource.
The Wigner-Yanase skew information can be used to derive physical implications related to the time-energy uncertainty relations and the estimation of quantum evolution speed \cite{Luo2}.
It has been experimentally confirmed that the robustness of coherence can quantify the advantage enabled by a quantum state in a phase discrimination task \cite{Zheng}, and it has been further demonstrated that the robustness of certain resources can be utilized in the calculation of operational advantages provided by resource states over free states in quantum state discrimination tasks within subchannels, while the weight of these resources can be similarly used in quantum state exclusion tasks within subchannels \cite{Takagi, Ducuara, Uola}.
This is also equally applicable to the resource theory of the incompatibility of quantum measurements \cite{Skrzypczyk1, Skrzypczyk2, Buscemi} and quantum channels \cite{Takagi2}.

We provide new frameworks and extended concepts for quantum resource theories by using new insights from relativity. We show that in the same context as the recent result that generalized measures of robustness for closed convex sets can exhibit practical operational advantages in quantum resource theories, the same results can be derived for measures of robustness introduced in the relative framework. Furthermore, based on the concept of relativity, we propose a new framework for quantifying the deficiency of quantum resources relative to the maximum resource states, and we present a geometric measure of resource deficiency that satisfies the conditions of this framework for both quantum coherence and quantum entanglement. Finally, we demonstrate that this geometric measure can also be used to clearly indicate operational disadvantages in subchannel discrimination. We believe that this extended quantum resource theory in the relative aspect will enable broader applications beyond robustness-based resource measures and geometric resource measures, as we discuss in the conclusion.

{\it Quantum resource theory in relative terms.--}
Let us consider the relativity of resources between quantum states from a resource-theoretic perspective.
We first characterize the relative resources between a given set of quantum devices (quantum states, quantum channels, quantum measurements, etc.) and any quantum devices not included in that set. Here we consider specific situations to clarify what is meant by a relative perspective.
The limitations of cost or technology make it impossible or difficult for us to prepare arbitrary quantum devices.
Conversely, some facilities (even if it costs money) allow the preparation of devices that, although limited, possess certain quantum potentials.

\begin{figure}
\includegraphics[width=3.0in]{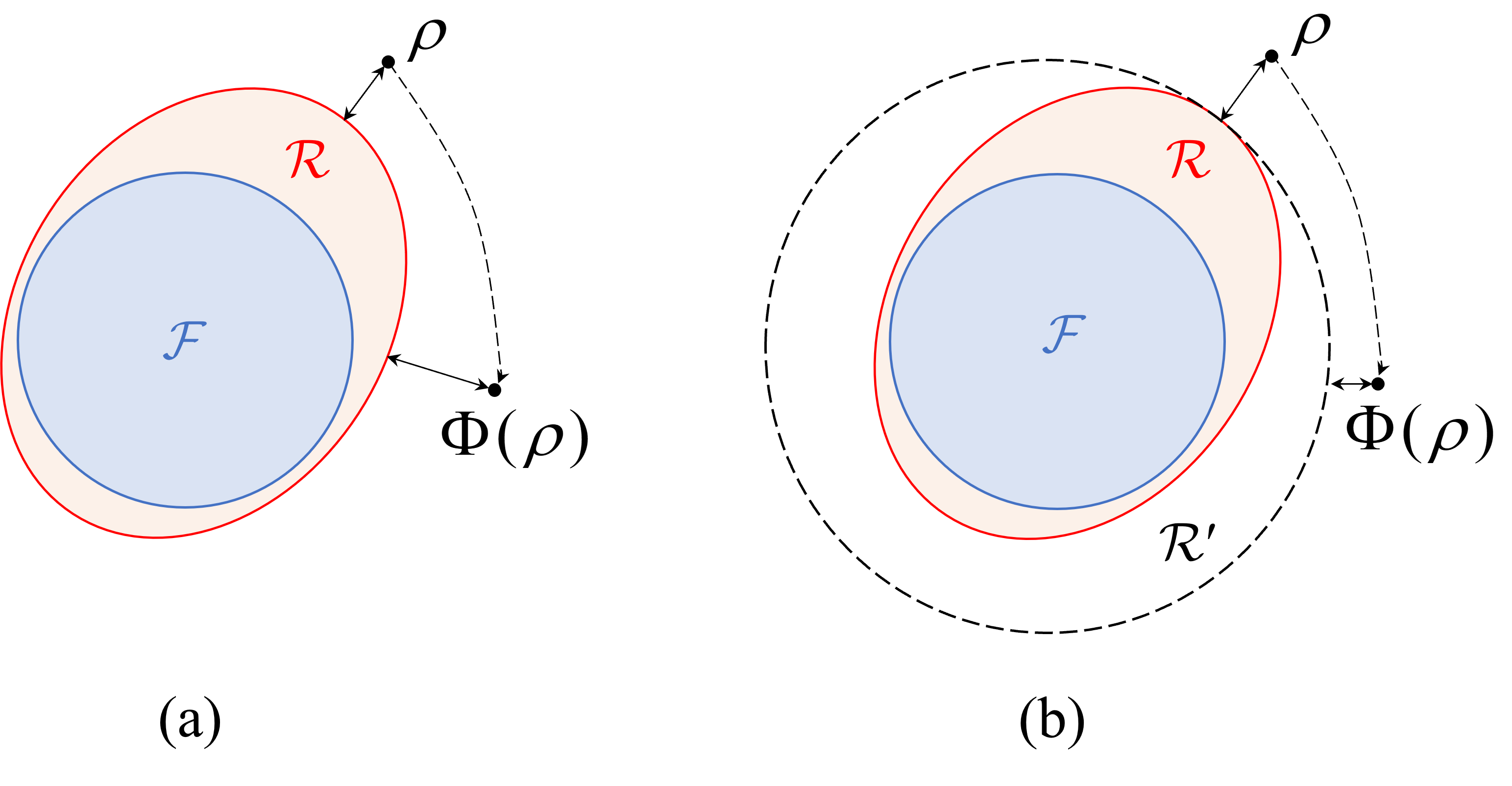}
\caption{(a) $\cF$ is the set of all free states and $\cR$ is the set of resource states that can be prepared through the current facility. Then, for some quantum states $\rho$ not in $\cR$, there is a free operation $\Phi$ that can increase the relative resources quantified for $\cR$. That is, monotonicity is not established. (b) We can find the minimum set that maintains monotonicity for free operations through the set $\cR'$ of all quantum states that do not exceed the maximum resource effect through the quantum states in the set $\cR$. The choice of the set $\cR'$ depends on the given quantum task and the corresponding measure of resources.}
\label{fig1}
\end{figure}

In this paper, we will restrict the subject of our main ideas to the representative quantum medium, the quantum states. Let $\cH$ be a $d$-dimensional Hilbert space and $\cD(\cH)$ be the set of density operators (quantum states) acting on $\cH$. Also, let $\cR(\cH)$ denote the set of all quantum states that we can prepare at cost or through special facilities (that must be the set of quantum states containing all free states).
Quantifying the relative advantages of resources of another quantum state that cannot be prepared with the current facility for set $\cR(\cH)$ serves as an example of the situations we consider.
As for our goal of quantifying relative resources, it is considered reasonable to quantify their relative resources through measures that satisfy faithful, monotonicity, and convexity, as in the original quantum resource theory.
This is because these are general and intuitive assumptions that reflect natural properties in the quantum physical setting. But we face a challenge : the form of a given set $\cR(\cH)$ may restrict or not be suitable for constructing measures that satisfy the above conditions.
If the given set $\cR(\cH)$ is not convex, it is difficult to derive a measure that satisfies the convexity.
In addition, if the resource graph of the set $\cR(\cH)$, quantified through a measure optimal for characterizing the resource properties associated with a given quantum mission, is distorted for the set of free states (see (a) of FIG.\ref{fig1}), it may not be possible to construct a measure that satisfies monotonicity for the free operation.
Thus we characterize the relative resources for an extended set that contains the given set $\cR(\cH)$. The key is how to extend the set to be adopted so that the characterized relative resources will be reasonably accepted. For this, we find one way to make sense of how to extend the set $\cR(\cH)$ from a particular situation.

For a quantum processing task, let $\rR$ be a suitable measure to characterize the resources in this task based on existing quantum resource theory.
If for two quantum states satisfying $\rR(\rho)<\rR(\rho')$ we judge that $\rho'$ inherent resources are superior to $\rho$ for this task.
It should be noted, however, that this numerical advantage may not be a comparison of numbers obtained from one identical strategy which is determined as a channel or measurement performed for the quantum task. In general, the measure $\rR$ is recognized as the maximum value of the resource effect that may be obtained from all strategies compared to the free states. It means that in some strategies, the effect given through $\rho$ may bring more advantages than the effect through $\rho'$. Quantifying resources through the maximum effect available in all strategies is a principle frequently adopted in existing resource theory, and likewise, it seems reasonable for us to apply the same principle in quantifying relative resources.
From the same principle, if the maximum effect (in all strategies that may be considered) due to any quantum state $\rho$ not in the set does not reach the maximum effect of the states in the set, we cannot judge that the resources of $\rho$ are dominant over the given set $\rR$, even if the strategy is not the same. Therefore, we quantify relative resources based on a set of all quantum states with a maximum effect smaller than the states in the given set $\rR$ in all practicable strategies (see (b) of FIG.\ref{fig1}), and the extended set is given as follows:
$\cR^\alpha_\rR(\cH) = \{\rho|\rR(\rho)\leq \alpha\}$ where $\alpha = \max\{\rR(\rho)|\rho \in \cR(\cH)\}$. This extended set is well suited to constructing a measure consistent with the resource theory given as faithfulness, monotonicity, and convexity. From this we present a framework of resource theory for quantifying relative resources.

Here, our purpose is to quantify the superior resource compared to all resource states in which the resource amount does not exceed $\alpha$ ($0\leq \alpha< \rR(\sigma_{\textmd{max}})$ where $\sigma_{\textmd{max}}$ is a maximally resource state) through a given measure $\rR$.
We refer to functions that can quantify these relative resources as $\alpha$-superiority measures of resource and the $\alpha$-superiority measures  $\rR_\alpha$ are defined as satisfying the following conditions:

($\alpha$-R1\label{r1}) faithful: $\rR_\alpha(\rho)\geq 0$, and $\rR_\alpha(\rho)=0$ if and only if $\rho \in \cR^\alpha_{\rR}(\cH)$ where $\cR^\alpha_{\rR}(\cH)$ is the set of all states $\rho$ satisfying $\rR(\rho)\leq \alpha$;

($\alpha$-R2a\label{r2}) monotonicity under any free operation $\Sigma$: $\rR_\alpha(\Sigma(\rho)) \leq \rR_\alpha(\rho)$ for any free operations $\Sigma$,

or ($\alpha$-R2b) monotonicity under selective measurement $\{K_n\}$ : $\sum_np_n\rR_\alpha(\sigma_n)\leq \rR_\alpha(\rho)$, where $\sigma_n = K_n\rho K_n^\dagger/p_n$ with $p_n = \tr{K_n\rho K_n^\dagger}$;

($\alpha$-R3\label{r3}) convexity: $\rR_\alpha(\sum_iq_i\rho_i) \leq \sum_iq_i\rR_\alpha(\rho_i)$. \\
Note that  (R2b, R3) imply (R2a). When (R1-R3) is satisfied, it is called a measure.
In addition, the inclusion relationship $\cF(\cH)\subset\cR_{\rR}^\alpha(\cH)\subset\cR^\beta_{\rR}(\cH)$ is established between the sets of quantum states given respectively according to any number $\alpha, \beta$ satisfying $0<\alpha<\beta$, and particularly, $\cF(\cH)=\cR_{\rR}^0(\cH)$.






{\it  Operational relative advantage in subchannel discrimination.--}
We already know that the measure of robustness represents an operational advantage for a closed convex set of quantum states in subchannel discrimination \cite{Takagi}. Here we also introduce our application to the relative resource theory presented by us as a particular situation of this result. For this, it is necessary to define a robustness measure for quantifying relative resources. First, from the existing generalized robustness measure $\rR_{r}$ such that
$$\rR_{r}(\rho) = \min_{\tau\in \cD(\cH)}\Big\{s\geq0\Big|\frac{\rho+s\tau}{1+s}\in \cF(\cH)\Big\},$$
when defined as a closed convex set $\cR^\alpha_{r}(\cH)$ of resource states with respect to $\alpha$, the relative robustness for that given set is defined as follows:
\begin{defi}\label{def1}
Given a state $\rho$, for the robustness measure $\rR_{r}$ and $0\leq \alpha< d-1$, we define the robustness of $\alpha$-superiority of $\rho$ as
\be\label{eq1}
\rR_{r,\alpha}(\rho) = \min_{\tau\in \cD(\cH)}\Big\{s\geq0\Big|\frac{\rho+s\tau}{1+s}\in \cR^\alpha_{r}(\cH)\Big\}
\ee
where $\cR^\alpha_{r}(\cH) = \{\sigma| \ \rR_r(\sigma)\leq \alpha \ \ \textmd{and}\ \ \sigma\in \cD(\cH)\}$.
\end{defi}

Then, the numerical relationship between $\rR_{r}$ and $\rR_{r,\alpha}$ proved in the following theorem can show that $\rR_{r,\alpha}$ is a measure of $\alpha$-superiority. (See the proof in Appendix A of the Supplemental Material \cite{kim}.)

\begin{thrm}\label{thm4}
For the robustness measure $\rR_{r}$ and $0\leq \alpha< d-1$, we have
\be\label{8}
\rR_{r,\alpha}(\rho)= \max\Big\{\frac{\rR_{r}(\rho)-\alpha}{1+\alpha},0\Big\}.
\ee
Therefore, $\rR_{r,\alpha}$ also satisfies (R1\ref{r1}-R3\ref{r3}).
\end{thrm}

Next, we add a more specific situation to discuss the relative advantages in subchannel discrimination. Apart from the robustness measure presented above, we present a definition of the relative robustness of arbitrary quantum states for a given single quantum state regardless of the choice of quantum resources. From this we show that the relative operational advantages of replacing a set of closed convex states with a single quantum state are given equally by the relative robustness between the single quantum states that we present.
The relative robustness of arbitrary quantum states $\rho$ for a given quantum state $\sigma$ is given as follows:
\be
\rR^+(\rho, \sigma) = \min_{\tau\in \cD(\cH)}\Big\{s\geq0\Big|\frac{\rho+s\tau}{1+s}=\sigma\Big\}
\ee
We refer to the subspace of $\cH$ projected through all eigenstates with nonzero eigenvalues for any $\rho$ as $\cH_\rho$.
If $\cH_\rho\nsubseteq\cH_\sigma$,  the above values cannot be determined, and in this case, we define as $\rR_{r}^+(\rho, \sigma)=\infty$.

To come back and complete our goal, we briefly explain the definition and process of subchannel discrimination. Let $\{\Psi_i\}$ denote a finite set of subchannels (completely positive trace-nonincreasing maps) that compose a completely positive trace-preserving (CPTP) map $\Lambda = \sum_i \Psi_i$. We consider the discrimination task which subchannel has been applied to the input state $\rho \in \cD(\cH)$, where we perform measurements $\{M_i\}$ on the output under the promise that only one of the subchannels within the set is realized. The success probability for subchannel discrimination for any given strategy $(\{\Psi_i\}, \{M_i\})$ is $P_{succ}(\{\Psi_i\}, \{M_i\}, \rho) = \sum_i\tr{M_i\Psi_i(\rho)}$.
Then, we have the following results on the maximum ratio of the success probability of subchannel discrimination through the two quantum states.
(See the proof in Appendix B of the Supplemental Material \cite{kim}.)

\begin{lem}\label{lem1}
When we define the set of all enforceable strategies $(\{\Psi_i\}, \{M_i\})$ as $\Omega$ in the subchannel quantum discrimination task, we obtain that
\be\label{eq1}
\max_{\Omega}\frac{P_{succ}(\{\Psi_i\}, \{M_i\}, \rho)}{P_{succ}(\{\Psi_i\}, \{M_i\}, \sigma)} = 1+ \rR^+(\rho, \sigma).
\ee
This is the maximum ratio between the probabilities of success through all the strategies in $\Omega$
(In this paper, only for $(\{\Psi_i\}, \{M_i\})\in \Omega$, the application range of the functions max and min will be directly simplified and expressed as $\Omega$).
\end{lem}

The relative operational advantages of two single quantum states have significant implications for the choice of strategies to implement the mission, regardless of the influence of any quantum resource. In other words, the optimal ratio of success probability for every choice of all viable strategies $(\{\Psi_i\}, \{M_i\})$ in subchannel discrimination is given by the relative robustness, where the relative advantages given do not represent the relative superiority of the resources of the quantum states themselves, but only the contextual advantages given in the choice of implementable strategies.
Furthermore, this can naturally be extended to the following results in relation to the closed convex resource state set presented earlier.

\begin{coro}\label{coro1}
Given a state $\rho$, for the robustness measure $\rR_{r}$ and $0\leq \alpha< d-1$, we can rewrite the $\alpha$-superiority of $\rho$ via robustness as
\bea
\rR_{r,\alpha}(\rho)= \min_{\sigma \in \cR^\alpha_{r}}\rR^+(\rho, \sigma).
\eea
So, from Eq. (\ref{eq1}) and Theorem \ref{thm4}, we have that
\bea
\min_{\sigma\in \cR^\alpha_{r}}\max_{\Omega}\frac{P_{succ}(\{\Psi_i\}, \{M_i\}, \rho)}{P_{succ}(\{\Psi_i\}, \{M_i\}, \sigma)} &=& 1+ \rR_{r,\alpha}(\rho)\\
 &=& \max\{\frac{1+\rR_{r}(\rho)}{1+\alpha},1\}.
\eea
\end{coro}

This result shows that for a closed convex set of resource states fixed by a given $\alpha$ in context, we can compute the operational relative advantages of arbitrary quantum states in subchannel discrimination as the robustness measure of $\alpha$-superiority. This clearly describes one particular situation about the result that operational advantages for any closed convex set can be quantified by robustness measure shown in \cite{Takagi}.
We also show that operational advantages in subchannel discrimination mission are decreasing in inverse proportion to $1+\alpha$ with increasing $\alpha$ specifying a set of resource states by providing a numerical relationship between the measure of $\alpha$-superiority and the original generalized robustness measure.


{\it Quantum resource theory of deficiency for the maximum resource states.--}
We introduced the theory of resources from a relative perspective through a special situation in the preceding section. Here we attempt to characterize the relative resources from a different perspective. In theory, quantum states with maximum resources can be used as the primary material that provides the greatest efficiency in various quantum communication and computational task, which has also been confirmed by several experiments. However, it is not always guaranteed that the quantum states we have prepared for various missions are maximum resource states.  Therefore, it can be very useful in practice to evaluate how inefficient a prepared quantum state is compared with maximum resource states. Additionally, considering that for most quantum resources such as entanglement, coherence, and magic, the set of maximum resource states is the set of pure states corresponding to the boundary of the convex whole set (which does not imply that all pure states are maximum resources), it is advantageous to propose a framework that inversely applies existing resource theory to assess and measure the deficiency of a given resource state. The reason is clarified by an example of a deficiency measure presented later.
From this, we present a new framework for measuring the degree of resource deficiency through conditions contrary to the fundamental properties of quantum resources in the existing quantum resource theory. The measure of resource deficiency for maximum resource states is defined by a function $\rD$, which satisfies the following conditions:

(D1) faithful: $\rD(\sigma)\geq 0$, and $\rD(\sigma)=0$ if and only if $\sigma \in \overline{\cR^{\max}}$ where $\overline{\cR^{\max}}$ is the set of all maximum resource states $\sigma$;

(D2a) (nondecreasing) monotonicity under any free operation $\Phi$: $\rD(\Phi(\rho)) \geq \rD(\rho)$,\\
or (D2b) monotonicity under selective measurement $\{K_n\}$ : $\sum_np_n\rD(\rho_n)\geq \rD(\rho)$, where $\rho_n = K_n\rho K_n^\dagger/p_n$ with $p_n = \tr{K_n\rho K_n^\dagger}$;


(D3) concavity: $\rD(\sum_iq_i\sigma_i) \geq \sum_iq_i\rD(\sigma_i)$. \\
These three conditions mean respectively that as the measured value increases, the deficiency compared to the maximum resource states also increases (from D1), the deficiency increases conversely through the reduction of resources via free operations (from D2a and D2b), and the convexity of the resource is consistent with the concavity of the deficiency (from D3).
One important aspect to consider is that the maximum resource set is usually comprised only of pure states. This implies that some measures, including the robustness-based measure, do not meet the conditions of this framework for quantifying relative deficiencies, as pure states are not attainable through any combination of quantum states. Therefore, we suggest here employing a geometric measure to quantify resource deficiencies.

\label{def3}
Given a state $\rho$, we define the geometric function for the maximum resource states in $\rho$,
\be\label{eq2}
\rD_g(\rho) = \min_{\sigma\in\overline{\cR^{\max}}}\Big\{1-\textmd{F}(\sigma,\rho)\Big\}
\ee
where the fidelity $\rF(\sigma,\rho) = \|\sqrt{\sigma}\sqrt{\rho}\|_1^2$ for two positive semidefinite operators $\sigma, \rho$.
In addition, considering that all maximum resource states are pure, $\rR_{\overline{\max}}$ can be expressed as $\rD_g(\rho) = \min_{\sigma\in\overline{\cR^{\max}}}\{1-\inner{\Pi_\sigma}{\rho}\}$
where $\Pi_\sigma$ is the projection operator over all eigen-states with nonzero eigenvalues for $\sigma$.

We next check that this geometric function $\rD_g$ is suitable for measuring resource deficiency for coherence and entanglement, respectively.
(See the proofs of the following two theorems in Appendix C of the Supplemental Material \cite{kim}.)
\begin{thrm}\label{thm5}
We define
\be\label{eq24}
\rD^C_g(\rho) = \min_{\sigma\in\overline{\cC^{\max}}}\Big\{1-\rF(\sigma,\rho)\Big\}
\ee
where $\overline{\cC^{\max}}$ is is the set of all maximally coherent states.
Then, $\rD^C_g$ is a measure of coherence deficiency.
\end{thrm}

\begin{thrm}\label{thm6}
We define
\be\label{eq29}
\rD^E_g(\rho) = \min_{\sigma\in\overline{\cE^{\max}}}\Big\{1-\rF(\sigma,\rho)\Big\}
\ee
where $\overline{\cE^{\max}}$ is is the set of all maximally entangled states.
Then, $\rD^E_g$ is a measure of entanglement deficiency.

\end{thrm}

{\it Operational disadvantage for maximum resources in subchannel discrimination.--}
Our final goal is to present a indicator of operational disadvantages of quantum states in subchannel discrimination and to establish a relationship between that indicator and resource deficiency. Before that, we first slightly modify the form of the indicator for the advantages presented in \cite{Takagi} in order to present a indicator for more reasonable disadvantages.  In order to represent the relative advantages of quantum states in subchannel quantum discrimination, as shown in equation (1), the maximum ratio of success probabilities generally given among all possible strategies is used as an indicator. However, since this only considers that the ratio of success probabilities is maximized, the strategy to maximize that ratio does not guarantee a sufficiently large probabilities of success for the quantum states to which it is executed. To secure this, we present an improved form of indicator to represent operational disadvantages, which is to calculate the relative ratio over only the plans that provide maximum success probability for the maximum resource state, not the percentage for all strategies. That is
\be\label{def34}
\max_{\sigma\in \cR^{\overline{max}}}\min_{\Omega_\sigma}\frac{P_{succ}(\{\Psi_i\}, \{M_i\}, \rho)}{P_{succ}(\{\Psi_i\}, \{M_i\},\sigma)}
\ee
where $\Omega_\sigma = \big\{(\{\Psi_i\}, \{M_i\})|P_{succ}(\{\Psi_i\}, \{M_i\}, \sigma) = 1\big\}.$
Under the strategy with the maximum success probability for the maximum resource state in Eq. (\ref{def34}), the minimum ratio between the success probabilities through two quantum states represents the relative disadvantage that the quantum state $\rho$ has for the maximum resource state $\sigma$, and from the maximum for all the maximum resource states in that ratio we can calculate the relative disadvantage for the maximum set of resources in the quantum state $\rho$. In other words, as the value increases, the relative disadvantage decreases, and as the value decreases, the degree of disadvantage increases. Moreover, by a strategy that provides the value of the disadvantage, we still guarantee the maximum success probability for the maximum resource state that give the maximum ratio.

In the indicator presented above, we only adopt strategies that provide the maximum success probability for the maximum resource states. And considering that the maximum resource states are generally pure states, we present the conditions for strategies that ensure that the maximum success probability is 1 and show that there are countless strategies that satisfy that conditions in Appendix D of the Supplemental Material \cite{kim}.

Furthermore, the following result shows that operational disadvantages due to quantum states of the newly proposed form in subchannel discrimination can be represented by surprisingly geometric measure of resource deficiency. (See the proof in Appendix E of the Supplemental Material \cite{kim}.)

\begin{thrm}\label{thm7}
For any $\rho\in \cD(\cH)$,
\be\label{36}
\max_{\sigma\in \cR^{\overline{max}}}\min_{\Omega_\sigma}\frac{P_{succ}(\{\Psi_i\}, \{M_i\}, \rho)}{P_{succ}(\{\Psi_i\}, \{M_i\},\sigma)} = 1-\rD_g(\rho).
\ee
\end{thrm}

This result shows that for a closed concave set of maximal resource states, the operational relative disadvantages of arbitrary quantum states in the identification of subchannels can be calculated as a geometric measure of resource deficiency.
This, contrary to the use of a robustness measure to quantify the operational advantages for an arbitrary closed convex set, explains that geometric measure can be used to quantify operational disadvantages, even if the robustness measure cannot be defined to quantify the resource deficiencies.
Also, for any quantum resource where all maximal resource states exist only in pure states, we can apply this result.
However, this conclusion does not apply equally to an arbitrary closed concave set as a relation determined from the special form of maximal resource states.

{\it Conclusion.--}
In this work we have dealt a broader perspective of quantum resource theory, where resources are relative to a specifically given set of resource states, rather than the traditional binary view of free states and resource states in existing quantum resource theories. This has led us to introduce a new framework for resource theory with two main aspects: resource $\alpha$-superiority and resource deficiency.

The first main result is that the generalized measures of robustness for certain convex sets of resources are satisfied in the framework of the resource theory of advantage, and it is also shown that the previous results of 'generalized measures of robustness exhibit operational advantages in subchannel discrimination' are also valid in the relative aspects of certain convex sets of resource states. In other words, the robustness measure of resource advantage that we proposed can be used to characterize the relative operational advantage in subchannel discrimination.

The second aspect we considered was quantifying the degree of resource deficiency for sets of maximally resourceful states, given that many existing quantum resource theories and experiments consider maximally resourceful states as the optimal quantum resources and the main problem is how close the prepared quantum states are to the maximum resource states. To address this, we proposed a new framework for resource theory that considers the physical principles employed in existing quantum resource theories in reverse order to quantify resource deficiency for sets of maximum resource states. We then proved that the geometric measure we proposed fulfills the conditions for quantifying resource deficiency with respect to quantum coherence and quantum entanglement. Finally, we showed that the operational disadvantage for sets of maximum resource states in the subchannel discrimination can be clearly characterized by our proposed geometric measure.

There are several natural questions and extensions that we leave for future work. For example, we can extend the resourceful utilization of quantum measurements \cite{Skrzypczyk2}, which has been considered under the subchannel discrimination, to a relative perspective, and we can consider generalizing the resourcefulness to other quantum resource theories where several measurements are allowed instead of the subchannel discrimination.

{\it Acknowledgments.--}This project is supported by the National Natural Science Foundation of China (Grants No. 12050410232 and No.12201555).


\begin{thebibliography}{99}%

\bibitem{Bennett} C. H. Bennett, G. Brassard, S. Popescu, B. Schumacher, J. A. Smolin, W. K. Wootters, 
    Phys. Rev. Lett. {\bf76}, 722 (1996)

\bibitem{Plenio} M. B. Plenio, and S. Virmani, 
 Quantum Inf. Comput. {\bf 7}, 1-51 (2007)

\bibitem{Regula} B. Regula, 
 J. Phys. A {\bf51}, 045303 (2018)


\bibitem{Horodecki1} P. Horodecki and A. Ekert, 
 Phys. Rev. Lett. {\bf 89}, 127902 (2002)


\bibitem{Schaetz} T. Schaetz, M. D. Barrett, D. Leibfried, J. Britton, J. Chiaverini, W. M. Itano, J. D. Jost, E. Knill, C. Langer, and D. J. Wineland, 
     Phys. Rev. Lett.  {\bf 94}, 010501 (2005)

\bibitem{Hou} J. Hou, and X. Qi, 
 Phys. Rev. A {\bf81}, 062351 (2010)

\bibitem{Roa} L. Roa, J. C. Retamal, and M. Alid-Vaccarezza, 
Phys. Rev. Lett. {\bf 107}, 080401 (2011).

\bibitem{Horodecki2} R. Horodecki, P. Horodecki, M. Horodecki, and K. Horodecki, 
Rev. Mod. Phys. {\bf 81}, 865 (2009)

\bibitem{Streltsov} A. Streltsov, G. Adesso, and M.B. Plenio, 
 Rev. Mod. Phys. {\bf 89}, 041003 (2017)

\bibitem{Bartlett} S.D. Bartlett, T. Rudolph, and R.W. Spekkens, 
 Rev. Mod. Phys. {\bf 79}, 555 (2007)


\bibitem{Gour} G. Gour, M. P. Muller, V. Narasimhachar, R. W. Spekkens, and N. Y. Halpern, 
     Phys. Rep. {\bf 583}, 1 (2015)


\bibitem{Luo} S. Luo, and S. Fu, 
 Phys. Rev. Lett. {\bf 106}, 120401 (2011)


\bibitem{Brunner} N. Brunner, D. Cavalcanti, S. Pironio, V. Scarani, and S. Wehner, 
 Rev. Mod. Phys. {\bf 86}, 419 (2014)


\bibitem{Weedbrook} C. Weedbrook, S. Pirandola, R. Garc\'{\i}a-Patr\'{o}n, N. J. Cerf, T. C. Ralph, J. H. Shapiro, and S. Lloyd, 
     Rev. Mod. Phys. {\bf 84}, 621 (2012)

\bibitem{Luo1} S. Luo, 
 Phys. Rev. A {\bf 77}, 042303 (2008)

\bibitem{Modi} K. Modi, A. Brodutch, H. Cable, T. Paterek, and V. Vedral,
Rev. Mod. Phys. {\bf 84}, 1655 (2012).


\bibitem{Mintert} F. Mintert, M. Ku\'{s}, and A. Buchleitner, 
 Phys. Rev. Lett. {\bf 95}, 260502 (2005)

\bibitem{Vicente} J. I. de Vicente, C. Spee, and B. Kraus, 
 Phys. Rev. Lett. {\bf 111}, 110502 (2013)


\bibitem{Hou1} J. Hou, L. Liu, and X. Qi, 
     Phys. Rev. A {\bf 105}, 032429 (2022)


\bibitem{Gour1} G. Gour, and A. Winter, 
 Phys. Rev. Lett. {\bf 123}, 150401 (2019)

\bibitem{Gour2} G. Gour, 
 IEEE Trans. Inf. Theory {\bf 65}, 5880-5904 (2019)

\bibitem{Saxena} G. Saxena, E. Chitambar, G. Gour, 
 Phys. Rev. Research {\bf 2}, 023298 (2020)




\bibitem{Chitambar} E. Chitambar and G. Gour, 
 Rev. Mod. Phys. {\bf 91}, 025001 (2019)


\bibitem{Vidal} G. Vidal, and R. Tarrach, 
 Phys. Rev. A {\bf 59}, 141 (1999)

\bibitem{Harrow} A. W. Harrow, and M. A. Nielsen, 
 Phys. Rev. A {\bf 68}, 012308 (2003)

\bibitem{Napoli} C. Napoli, T. R. Bromley, M. Cianciaruso, M. Piani, N. Johnston, and G. Adesso, 
    Phys. Rev. Lett. {\bf 116}, 150502 (2016).


\bibitem{Vedral} V. Vedral, M. B. Plenio, M. A. Rippin, and P. L. Knight, 
 Phys. Rev. Lett. {\bf 78}, 2275 (1997)

\bibitem{Baumgratz} T. Baumgratz, M. Cramer, and M. B. Plenio, 
Phys. Rev. Lett. {\bf 113}, 140401 (2014).

\bibitem{Renyi} A. R\'{e}nyi, {\it 
Volume 1: Contributions to the Theory of Statistics}
    (University of California Press, Berkeley, 1961)


\bibitem{Eisert} J. Eisert, K. Audenaert, and M. B. Plenio, 
 J. Phys. A {\bf36}, 5605 (2003)

\bibitem{Winter} A. Winter and D. Yang, 
Phys. Rev. Lett. {\bf 116}, 120404 (2016).

\bibitem{Bravyi} S. Bravyi, and A. Kitaev, 
 Phys. Rev. A  {\bf71}, 022316 (2005)

\bibitem{Elitzur}  A. C. Elitzur, S. Popescu, and D. Rohrlich, 
 Phys. Lett. A {\bf162}, 25-28 (1992)

\bibitem{Lewenstein}  M. Lewenstein, and A. Sanpera, 
 Phys. Rev. Lett. {\bf80}, 2261 (1998)

\bibitem{Skrzypczyk}  P. Skrzypczyk, M. Navascu\'{e}s, and D. Cavalcanti, 
 Phys. Rev. Lett. {\bf112}, 180404 (2014)


\bibitem{Luo2} S. Luo, 
 Phys. Rev. Lett. {\bf91}, 180403 (2003)

\bibitem{Zheng} W. Zheng, Z. Ma, H. Wang, S. M. Fei, and X. Peng, 
     Phys. Rev. Lett. {\bf120}, 230504 (2018)


\bibitem{Takagi} R. Takagi, B. Regula, K. Bu, Z.-W. Liu, and G. Adesso, 
     Phys. Rev. Lett. {\bf122}, 140402 (2019)


\bibitem{Ducuara} A. F. Ducuara, and P. Skrzypczyk, 
     Phys. Rev. Lett. {\bf125}, 110401 (2020)

\bibitem{Uola} R. Uola, T. Bullock, T. Kraft, J.-P. Pellonp\"{a}\"{a}, and N. Brunner, 
 Phys. Rev. Lett. {\bf125}, 110402 (2020)

\bibitem{Skrzypczyk1} P. Skrzypczyk, and I. \u{S}upi\'{c}, 
     Phys. Rev. Lett. {\bf122}, 130403 (2019)

\bibitem{Skrzypczyk2} P. Skrzypczyk, and N. Linden, 
Phys. Rev. Lett. {\bf122}, 140403 (2019)

\bibitem{Buscemi} F. Buscemi, E. Chitambar, and W. Zhou, 
 Phys. Rev. Lett. {\bf124}, 120401 (2020)

\bibitem{Takagi2} R. Takagi, K. Wang, and M. Hayashi, 
 Phys. Rev. Lett. {\bf124}, 120502 (2020)

\bibitem{kim} See the Supplemental Material for detailed discussions and proofs of the results in the main text, which includes Refs. [38, 46-48].

\bibitem{Regula} B. Regula, 
J. Phys. A {\bf 51}, 045303 (2018).


\bibitem{Brandao} F. G. S. L. Brand\~{a}o, 
Phys. Rev. A 72, 022310 (2005).

\bibitem{Watrous}  J. Watrous, {\it Theory of Quantum Information} (Institute for Quantum Computing, University of Waterloo, Waterloo, Canada, 2008).


\end{thebibliography}

\begin{thebibliography}{99}%


\bibitem{Regula} B. Regula, 
J. Phys. A {\bf 51}, 045303 (2018).



\bibitem{Brandao} F. G. S. L. Brand\~{a}o, 
Phys. Rev. A 72, 022310 (2005).

\bibitem{Takagi} R. Takagi, B. Regula, K. Bu, Z.-W. Liu, and G. Adesso, 
     Phys. Rev. Lett. {\bf122}, 140402 (2019)


\bibitem{Watrous}  J. Watrous, {\it Theory of Quantum Information} (Institute for Quantum Computing, University of Waterloo, Waterloo, Canada, 2008).



\end{thebibliography}

%

\appendix

\newpage

\section{Supplemental Material:\\
Relative Quantum Resource Theory and Operational Applications in Subchannel Discrimination}

\begin{center}
{\bf Appendix A: Proof of Theorem 2}
\end{center}

To prove that $\rR_{r,\alpha}(\rho)=\frac{\rR_{r}(\rho)-\alpha}{1+\alpha}$ if $\rR_{D}(\rho) > \alpha$, we first prove $\rR_{r,\alpha}(\rho)\leq\frac{\rR_{r}(\rho)-\alpha}{1+\alpha}$. Then,
there is a state $\tau'$ satisfying $\frac{\rho+\rR_{r}(\rho)\tau'}{1+\rR_{r}(\rho)}\in\cF(\cH)$, and then, from the definition of $\rR_{r}$ and
\be\label{9}
\frac{\rho+\rR_{r}(\rho)\tau'}{1+\rR_{r}(\rho)} = \frac{\frac{\rho+(\frac{\rR_{r}(\rho)-\alpha}{1+\alpha})\tau'}{1+(\frac{\rR_{r}(\rho)-\alpha}{1+\alpha})}+\alpha\tau'}{1+\alpha}
\ee
we get that $\rR_{r}(\rho') \leq \alpha$ where $\rho' = \frac{\rho+(\frac{\rR_{r}(\rho)-\alpha}{1+\alpha})\tau'}{1+(\frac{\rR_{r}(\rho)-\alpha}{1+\alpha})}$. This means that $\rho' \in\cR^\alpha_{\rR_{r}}(\cH)$ from the definition of $\cR^\alpha_{\rR_{r}}(\cH)$.
It also implies that
\be\label{10}
\rR_{r,\alpha}(\rho)\leq\frac{\rR_{r}(\rho)-\alpha}{1+\alpha}
\ee
by the minimum definition of $\rR_{r,\alpha}$.

Conversely, to prove $\rR_{r,\alpha}(\rho)\geq\frac{\rR_{r}(\rho)-\alpha}{1+\alpha}$, let $\tau''$ be the state that satisfies $\frac{\rho+\rR_{r,\alpha}(\rho)\tau''}{1+\rR_{r,\alpha}(\rho)}\in\cR^\alpha_{\rR_{r}}(\cH)$. This implies that $\rR_{r}(\rho'')\leq\alpha$ where $\rho'' = \frac{\rho+\rR_{r,\alpha}(\rho)\tau''}{1+\rR_{r,\alpha}(\rho)}$, and also means that there is $\tau'''$ satisfying
\be\label{11}
\frac{\rho''+\rR_{r}(\rho'')\tau'''}{1+\rR_{r}(\rho'')}\in \cF(\cH).
\ee
And this is equivalent to
\be\label{12}
\frac{\rho+\Big\{\rR_{r,\alpha}(\rho)\tau''+\big(1+\rR_{r,\alpha}(\rho)\big)\rR_{r}(\rho'')\tau'''\Big\}}{1+\Big\{\rR_{r,\alpha}(\rho)+\big(1+\rR_{r,\alpha}(\rho)\big)\rR_{r}(\rho'')\Big\}}\in \cF(\cH).
\ee
and from the definition of $\rR_{r}$ we get that
\bea
\rR_{r,\alpha}(\rho)+\big(1+\rR_{r,\alpha}(\rho)\big)\rR_{r}(\rho'')\geq \rR_{r}(\rho).
\eea
Also, due to the inequality $\rR_{r}(\rho'')\leq\alpha$, this leads to
\be\label{13}
\rR_{r,\alpha}(\rho)\geq\frac{\rR_{r}(\rho)-\alpha}{1+\alpha}.
\ee

In addition, if $\rR_{r}(\rho) \leq \alpha$, since $\rho\in \cR^\alpha_{r}(\cH)$, $\rR_{r,\alpha}(\rho)= 0$ is directly obtained.
Therefore, for any state $\rho\in\cD(\cH)$, Eq. (2) in main text is established.

\begin{center}
{\bf Appendix B: Proof of Lemma 3}
\end{center}

For the relative robustness
\be
\rR^+(\rho, \sigma) = \min_{\tau\in \cD(\cH)}\Big\{s\geq0\Big|\frac{\rho+s\tau}{1+s}=\sigma\Big\},
\ee
it is also obtained as the optimal value of the convex optimization problem between two single quantum states, as a reduced form of the convex optimization problem of the generalized robustness measure (see Refs. \cite{Regula, Brandao, Takagi}), \emph{i.e.},
\begin{eqnarray}
\textmd{maximize} &&\quad \tr{\rho X}-1\qquad\label{17}\\
\textmd{subject to} &&\quad X\geq 0,\label{18}\\
&&\quad \tr{\sigma X} \leq 1 \label{19}.
\end{eqnarray}

To prove Lemma 3, we similarly use the proof of Theorem 2 in \cite{Takagi}.
We first recall the definition of relative robustness $\rR^+$. If $\rR^+(\rho, \sigma)< \infty$, there is a quantum state $\tau\in \cD(\cH)$ such that $\rho+\rR^+(\rho, \sigma)\tau = [1+\rR^+(\rho, \sigma)]\sigma$, it implies that $\sum_i\tr{M_i\Psi_i(\rho)}\leq[1+\rR^+(\rho, \sigma)]\sum_i\tr{M_i\Psi_i(\sigma)}$. From this, for any strategy $(\{\Psi_i\}, \{M_i\})$,
\bea
\frac{P_{succ}(\{\Psi_i\}, \{M_i\}, \rho)}{P_{succ}(\{\Psi_i\}, \{M_i\}, \sigma)} &=& \frac{\sum_i\tr{M_i\Psi_i(\rho)}}{\sum_i\tr{M_i\Psi_i(\sigma)}}\\
&\leq& 1+\rR^+(\rho, \sigma).
\eea
Next, we show that there is a strategy $(\{\Psi_i\}, \{M_i\})$ that satisfies the above equation. Let $X$ is an operator satisfying Eqs. (\ref{18}) and (\ref{19}) with the spectral decomposition as $X = \sum_i\lambda_i \out{\upsilon_i}{\upsilon_i} \ (\forall \ \lambda_i\geq0)$.
We then select the set of unitary operations $\{U_i\}$ that satisfy $\sum_i U_i\out{\upsilon_j}{\upsilon_j}U_i^\dagger = \id$ for all $j$. Now, consider a strategy $(\{\Psi_i\}, \{M_i\})$ defined by $\Psi_i(\cdot) = p_i U_i(\cdot) U_i^\dagger$ with a probability distribution $(p_i)$ and $M_i = U_iX U_i^\dagger\big/\tr{X}$ (then we can easily confirm that $\{M_i\}$ is a POVM). From the choice of the strategy $(\{\Psi_i\}, \{M_i\})$ we have that
\bea
\frac{P_{succ}(\{\Psi_i\}, \{M_i\}, \rho)}{P_{succ}(\{\Psi_i\}, \{M_i\}, \sigma)} &=& \frac{\sum_i p_i\tr{U_iX U_i^\dagger U_i\rho U_i^\dagger}}{\sum_i p_i\tr{U_iX U_i^\dagger U_i\sigma U_i^\dagger}}\\
&=& \frac{\tr{X\rho}}{\tr{X\sigma}}\geq \tr{X\rho}.
\eea
We find that for the optimized $X$ satisfying (\ref{17}-\ref{19}), $\tr{X\sigma}=1$ and that $X$ realizes $$\frac{P_{succ}(\{\Psi_i\}, \{M_i\}, \rho)}{P_{succ}(\{\Psi_i\}, \{M_i\}, \sigma)} = 1+ \rR^+(\rho, \sigma).$$

If $\cH_\rho\nsubseteq\cH_\sigma$, \emph{i.e.},$\rR_{r}^+(\rho, \sigma)=\infty$, there is a set of mutually orthogonal vectors $\{\ket{\varphi_i}\}_{i\in\Sigma}\ (1\leq|\Sigma|< d)$ such that $\iinner{\varphi_i|\rho}{\varphi_i}>0$ but $\iinner{\varphi_i|\sigma}{\varphi_i}=0$ for all $i$. We consider a strategy $(\{\Psi_i\}, \{M_i\})$ that the subchannels are defined by $\Psi_i(\cdot) = p_iU_i(\cdot)U_i$ where $U_i$ are unitary operations with their respective invariant vectors $\ket{\varphi_i}$ if $i\in\Sigma$ and by product of any POVM $\Lambda_i$ and $p_i$ if $i\notin \Sigma$, \emph{i.e.}, $\Psi_i(\cdot) = p_i\Lambda_i(\cdot)$, and the measurement $\{M_i\}$ is defined by $M_i = \out{\varphi_i}{\varphi_i}$ if $i\in\Sigma$ and $\sum_{j\notin\Sigma}M_j = \id - \sum_{i\in\Sigma}M_i$. Then, for $i\in\Sigma$, we have $\tr{M_i\Psi_i(\rho)}>0$ and $\tr{M_i\Psi_i(\sigma)}=0$. Therefore, the rate of the success probability can be considered infinite if $\sum_{i\in\Sigma}p_i=1$.

\begin{center}
{\bf Appendix C: Proofs of Theorem 5 and Theorem 6}
\end{center}

The geometric function $\rD_g$ we defined in main text always holds the conditions of (D1) and (D3) for any quantum resource.
First, we check about (D1). The fidelity between all quantum states is less than or equal to $1$, so $\rD_g(\rho)\geq 0$. And if $\rho$ is in the maximum resource state, from that definition, $\rD_g(\rho)= 0$, on the contrary, if $\rD_g(\rho)= 0$, there exists $\sigma$ that satisfies $\rF(\rho,\sigma)=1$, which means $\rho=\sigma$, where $\rho$ is the maximum resource state.
After this, for convenience, we define the maximum fidelity under the maximum resource states as follows:
\be
\rF_{\cR}(\rho) = \max_{\sigma\in\overline{\cR^{\max}}}\rF(\sigma,\rho)
\ee
Next, from the convexity of the maximum value function $\rF_{\cR}$,
\begin{eqnarray}\label{28}
\rD_g(\sum_iq_i\sigma_i) &=& 1-\rF_{\cR}(\sum_iq_i\sigma_i)\nonumber\\
 &\geq& 1-\sum_iq_i\rF_{\cR}(\sigma_i) = \sum_iq_i\rD_g(\sigma_i),\ \ \
\end{eqnarray}
therefore, (D3) is satisfied.

{\it Proofs of Theorem 5.--}
We only need to prove that it satisfies (D2b) to confirm that $\rD^C_g$ is the measure of deficiency. By the arbitrary maximally coherent states $\out{\psi}{\psi}$ are in the form of $\ket{\psi} = \frac{1}{\sqrt{d}}\sum_{i}e^{i\theta_i}\ket{i}$, they are derived as follows:
\be
\rF_{\cC}(\rho) = \max_{\sigma\in\overline{\cC^{\max}}}\rF(\sigma,\rho)= \frac{1}{d}\max_{\{\theta_i\}_i}\{\sum_{i,j}e^{i\theta_{ij}}\rho_{ij}\}
\ee
where $\rho_{ij} = \iinner{i}{\rho|j}$ and $e^{i\theta_{ij}} = e^{i(\theta_j-\theta_i)}$. It implies that $\rF_{\cC}(\rho) \leq \frac{1}{d}\sum_{i,j}|\rho_{ij}|$, and the equation holds if $\rho$ is pure.
Therefore, for quantum coherence, we have
\be\label{eq3}
\rD^C_g(\rho) \geq 1-\frac{\sum_{i,j}|\rho_{ij}|}{d}
\ee
and if $\rho$ is pure, we have
\be\label{eq3}
\rD^C_g(\rho) = 1-\frac{\sum_{i,j}|\rho_{ij}|}{d}.
\ee

Let $\Phi_\pi$ is an incoherent operation acted on by a series of permutation matrices $\{P_{\pi_n}\}$, \emph{e.i.}, $\Phi_\pi(\rho) = \sum_np_nP_{\pi_n}\rho P_{\pi_n}^\dagger$.
Then, there is a set $\{\theta^{(n)}_i\}_i$ that reach the maximum of $\rF_{\cC}(\rho_n)$ with $\rho_n = P_{\pi_n}\rho P_{\pi_n}^\dagger$, it implies that
\begin{eqnarray}\label{29}
\sum_np_n\rF_{\cC}(\rho_n) &=& \sum_np_n\sum_{i,j}e^{i\theta^{(n)}_{ij}} \rho_{\pi^{-1}_n(i)\pi^{-1}_n(j)}\nonumber\\
&\leq&  \sum_{i,j}e^{i\theta_{ij}}\rho_{ij} =\rF_{\cC}(\rho)
\end{eqnarray}
because $\sum_{i,j}e^{i\theta^{(n)}_{ij}} \rho_{\pi^{-1}_n(i)\pi^{-1}_n(j)}\leq \sum_{i,j}e^{i\theta_{ij}}\rho_{ij}$ for any $n$.
Hereby, we have that $\rD^C_g(\rho) \leq \sum_np_n\rD^C_g(P_{\pi_n}\rho P_{\pi_n}^\dagger).$

Next, for any incoherent operation $\Phi$, acting as $\Phi(\rho) = \sum_n K_n\rho K_n^\dagger$, let $k^{(n)}_j\ (j = 1,2,\cdots,d)$ be the nonzero element at the $j$th column of $K_n$ (if there is no nonzero element in the $j$th column, then $k^{(n)}_j=0$). Suppose $k^{(n)}_j$ locates the $f_n(j)$th row. Here, $f_n(j)$ is a function that maps $\{2,\cdots,d\}$ to $\{1,2,\cdots,d\}$ with the property that $1 \leq f_n(j) \leq j$. Let $\delta_{s,t} = 1\ (\textmd{if}\ s=t)$ or $0\ (\textmd{if}\ s\neq t)$.
Then there is a permutation $\pi_n$ such that
\bea
K_n&& \ = \\ P_{\pi_n}&&\small{\left(
                          \begin{array}{ccccc}
                            k_1^{(n)} & \delta_{1,f_n(2)}k_2^{(n)} & \cdots & \delta_{1,f_n(d-1)}k_{d-1}^{(n)} & \delta_{1,f_n(d)}k_d^{(n)} \\
                            & & & &\\
                            0 & \delta_{2,f_n(2)}k_2^{(n)} & \cdots & \delta_{2,f_n(d-1)}k_{d-1}^{(n)} & \delta_{2,f_n(d)}k_d^{(n)} \\
                            & & & &\\
                            0 & 0 & \cdots & \delta_{3,f_n(d-1)}k_{d-1}^{(n)} & \delta_{3,f_n(d)}k_d^{(n)} \\
                            & & & &\\
                            \vdots & \vdots & \ddots  & \vdots & \vdots \\
                            & & & &\\
                            0 & 0 & \cdots & 0 & \delta_{d,f_n(d)}k_d^{(n)} \\
                          \end{array}
                        \right)}.
\eea
From $\sum_n K_n^\dagger K_n = \id$, we get that
\bea\label{16}
\left\{
                          \begin{array}{l}
                          \sum_n |k_j^{(n)}|^2 = 1 \quad (j=1,2,\cdots,d),\\
                          \\
                          \sum_n \overline{k_1^{(n)}}\delta_{1,f_n(j)}k_j^{(n)} = 0 \quad (j=2,\cdots,d),\\
                          \\
                          \sum_n \sum_l \overline{k_i^{(n)}}k_j^{(n)}\delta_{l,f_n(i)}\delta_{l,f_n(j)} = 0 \\
                          \end{array}
                        \right.\\
                         \qquad \quad  (2\leq i<j\leq d \ \ \textrm{and} \ \ l=1,2,\cdots,i).
\eea

We can see from the definition of maximum fidelity $\rF$ that, for any $n$, $\rF$ increases when all $k^{(n)}_i$ in the matrix $K_n$ are placed in different rows. Therefore, we can prove the following without any loss, assuming that all $k^{(n)}_i$ are arranged in different rows each other.

There are sets $\{\theta^{(n)}_i\}$ and $\{\theta_i\}$ that allow us to acquire $\rF_{\cC}(\rho_n)$ and $\rF_{\cC}(\rho)$, \emph{i.e.}, $\rF_{\cC}(\rho_n) = \iinner{\phi_n}{\rho_n|\phi_n}$ and $\rF_{\cC}(\rho) = \iinner{\psi}{\rho|\psi}$ with
$\ket{\phi_n} = \frac{1}{\sqrt{d}}\sum_{i}e^{i\theta^{(n)}_i}\ket{i}$ and $\ket{\psi} = \frac{1}{\sqrt{d}}\sum_{i}e^{i\theta_i}\ket{i}$.
We define that $e^{i\theta'^{(n)}_{ij}} = e^{-i\theta_{ij}}e^{i\theta^{(n)}_{ij}}$ where $\theta_{ij} = \theta_{j}-\theta_{i}$ and $\theta^{(n)}_{ij} = \theta^{(n)}_{j}-\theta^{(n)}_{i}$.
Then, for any $i,j$, we have
\begin{eqnarray}\label{30}
\sum_n \textmd{re}\big\{e^{i\theta'^{(n)}_{ij}}k^{(n)}_{ij}\big\} &\leq& \sum_n |k^{(n)}_{ij}|\nonumber\\
 &\leq& \frac{\sum_n \big\{|k^{(n)}_i|^2+|k^{(n)}_j|^2\big\}}{2} = 1\ \
\end{eqnarray}
where $k^{(n)}_{ij}= k^{(n)}_i\overline{k^{(n)}_j}$, and it implies that
\bea\label{pf1}
\sum_np_n\rF_{\cC}(\rho_n) &\leq& \frac{1}{d}\sum_n \sum_{i,j}e^{i\theta^{(n)}_{ij}}k^{(n)}_{ij}\rho_{ij}\\
 &=&  \frac{1}{d}\sum_{i,j}\big(\sum_ne^{i\theta'^{(n)}_{ij}}k^{(n)}_{ij}\big)e^{i\theta_{ij}}\rho_{ij}\\
 &\leq& \frac{1}{d}\sum_{i,j}e^{i\theta_{ij}}\rho_{ij} = \rF_{\cC}(\rho).
\eea
Ineq. (\ref{29}) and our assumption induce the first inequality, and the last inequality is obtained through the fact that the Ineq. (\ref{30}) and $\rF_{\cC}$ is the maximum of the fidelity under all $\{\theta_i\}_i$.
Therefore, we have
\bea\label{pf2}
\sum_np_n\rD^C_g(\rho_n) &=& 1- \sum_np_n\rF_{\cC}(\rho_n)\\
 &\geq&  1- \rF_{\cC}(\rho) = \rD^C_g(\rho).
\eea

{\it Proof of Theorem 6.--} As in the case of coherence, this requires only a proof for (D2b) to confirm that $\rD^E_g$ is a measure of deficiency.
By the arbitrary maximally entangled states $\sigma =\out{\phi}{\phi}$ are in the form of $\ket{\phi} = \frac{1}{\sqrt{d}}\sum_{i}\ket{\phi_i}_A\ket{\phi_i}_B$ where $\{\ket{\phi_i}_A\}_i$ and $\{\ket{\phi_i}_B\}_i$ are sets of orthogonal states in systems $A$ and $B$, respectively, they are derived as follows:
\be
\rF_{\cE}(\rho) = \max_{\sigma\in\overline{\cE^{\max}}}\rF(\sigma,\rho)= \frac{1}{d}\max_{\scriptsize\begin{array}{c}
              \{\ket{\phi_i}_A\}_i \\
              \{\ket{\phi_i}_B\}_i
            \end{array}}\{\sum_{i,j}\rho^{(\phi)}_{iijj}\}
\ee
where $\rho^{(\phi)}_{efgh} = \iinner{\phi_{e}|_A\langle\phi_{f}|_B\rho}{\phi_{g}\rangle_A|\phi_h}_B$. If $\rho$ is pure, \emph{i.e.}, $\rho=\out{\psi}{\psi}$ where $\ket{\psi} = \sum_iq_i\ket{\psi_i}_A\ket{\psi_i}_B$ with a positive real $q_i$ for every $i$, we have
\be\label{eq3}
\rD^E_g(\rho) = 1-\frac{\sum_{i,j}q_iq_j}{d}.
\ee

Let $\Phi_U$ is a local unitary operation acted on by unitary matrices $U_A$ and $U_B$, \emph{e.i.}, $\Phi_U(\rho) = U_A\ot U_B\rho U_A^\dagger\ot U_B^\dagger$.
There are sets $\{\ket{\phi_i}_A\}_i$ and $\{\ket{\phi_i}_B\}_i$ that reach the maximum of $\rF_{\cE}(\rho)$, then
\be\label{33} \rD^E_g(\rho)=\rD^E_g\big[\Phi_U(\rho)\big]
\ee
is induced through
$$\rF_{\cE}\big[\Phi_U(\rho)\big] = \frac{1}{d}\sum_{i,j}\iinner{\phi'_{i}|_A\langle\phi'_{i}|_B\Phi_U(\rho)}{\phi'_{j}\rangle_A|\phi'_j}_B$$
 from the definition of $\rF_{\cE}$, where  $\ket{\phi'_i}_A = U_A\ket{\phi_i}_A$ and $\ket{\phi'_i}_B =U_B\ket{\phi_i}_B$ for any $i$.

Next, for any local operation $\Phi_{A}\ot \Phi_{B}$, acting as
$$\Phi_{A}\ot \Phi_{B}(\rho) = \sum_{n,m} p_{n,m}\rho_{n,m}$$
where $p_{n,m} = \tr{K^{(A)}_{n}\ot K^{(B)}_{m}\rho (K^{(A)}_{n})^\dagger\ot (K^{(B)}_{m})^\dagger}$ and $\rho_{n,m} = \big\{K^{(A)}_{n}\ot K^{(B)}_{m}\rho (K^{(A)}_{n})^\dagger\ot (K^{(B)}_{m})^\dagger\big\}/p_{n,m}$, we prove that $\sum_{n,m} p_{n,m}\rD^E_g(\rho_{n,m})\geq \rD^E_g(\rho)$.

To do this, we first consider local operation on a single system $\Phi_{A}\otimes \mathbb{I}_B$.
There are maximally entangled states $\sigma_n = \out{\phi^{(n)}}{\phi^{(n)}}$ for each $n$ with $\ket{\phi^{(n)}} = \frac{1}{\sqrt{d}}\sum_i\ket{\phi^{(n)}_i}_A\ot \ket{\phi^{(n)}_i}_B$, such that
\bea
\sum_{n} p_{n}\rF_{\cE}(\rho_{n}) = \sum_{n}p_{n}\iinner{\phi^{(n)}|\rho_{n}}{\phi^{(n)}}
\eea
 where $p_{n} = \tr{K^{(A)}_{n}\ot \id_{B}\rho (K^{(A)}_{n})^\dagger\ot \id_{B}}$ and $\rho_{n} = \big\{K^{(A)}_{n}\ot \id_{B}\rho (K^{(A)}_{n})^\dagger\ot \id_{B}\big\}/p_{n}$.
For all $n$, let $U^{(A)}_n\ot U^{(B)}_n$ be the unitary operator that satisfy $U^{(A)}_n\ket{\phi^{(n)}_i}_A = \ket{\phi_i}_A \ (i=1,2,\cdots, d)$ and $U^{(B)}_n\ket{\phi^{(n)}_j}_B = \ket{\phi_j}_B \ (j=1,2,\cdots, d)$, respectively, where $\ket{\phi} = \frac{1}{\sqrt{d}}\sum_i \ket{\phi_{i}}_A\ket{\phi_{i}}_B$ is the maximally entangled state that has the maximum fidelity with $\rho$, \emph{i.e.}, $\rF_{\cE}(\rho) = \iinner{\phi|\rho}{\phi}$. We consider here quantum states $\rho'_{n} = U^{(A)}_n\ot U^{(B)}_n\rho_{n}(U^{(A)}_n)^\dagger\ot (U^{(B)}_n)^\dagger$ for any $n$,
then $\sum_{n} p_{n}\rF_{\cE}(\rho_{n}) = \sum_{n} p_{n}\rF_{\cE}(\rho'_{n})$ is obtained from Eq. (\ref{33}). Hereby, we have
\be\label{34}
\sum_{n} p_{n}\rF_{\cE}(\rho_{n}) = \iinner{\phi|(\sum_{n} p_{n}\rho'_{n})}{\phi}.
\ee

Let $k^{(n)}_{A,ij} = \iinner{\phi_{i}|_A U^{(A)}_nK^{(A)}_{n}}{\phi_{j}}_A$, then from $\sum_n (K^{(A)}_{n})^\dagger K^{(A)}_{n} = \id_A$, we get $\sum_l \sum_n \overline{k^{(n)}_{A,li}}k^{(n)}_{A,lj} = \delta_{ij}$.
And, if unitary operator $U_{A}$ satisfies $(\sum_n|k^{(n)}_{A,ei}|^2)_e\succ (|u_{A,ei}|^2)_e$ for all $i$, here $\succ$ is the majorization relation between vectors \cite{Watrous} and $u_{A,ei} = \iinner{\phi_{e}|_A U_{A}}{\phi_{i}}_A$, the following inequality is established from the Cauchy-Schwarz Inequality and property of Schur-concave function for every $i,j$:
\bea\label{35}
\sum_{e,f}\sum_n|k^{(n)}_{A,ei}k^{(n)}_{A,fj}| &\leq& \sum_{e,f}\sqrt{\sum_n|k^{(n)}_{A,ei}|^2}\sqrt{\sum_n|k^{(n)}_{A,fj}|^2}\\
 &\leq& \sum_{e,f}|u_{A,ei}u_{A,fj}|.
\eea
We already know that there are infinitely many unitary operators that satisfy the above majorization relation, and we can then select an appropriate unitary operator $U_{A}$ that satisfies the following inequality for each $i,j$ according to the quantum state $\rho$:
\bea\label{36}
\sum_{i,j}\sum_{e,f}\big(\sum_{n}\overline{k^{(n)}_{A,ei}}k^{(n)}_{A,fj}\big)\rho^{(\phi)}_{eifj}\leq \sum_{i,j}\sum_{e,f}\overline{u_{A,ei}}u_{A,fj}\rho^{(\phi)}_{eifj},
\eea
and also obtain the following results from the definition of the maximum value of $\rF_{\cE}$:
\bea\label{37}
\sum_{n} p_{n}\rF_{\cE}(\rho_{n})&=& \frac{1}{d}\sum_{i,j}\sum_{e,f}\big(\sum_{n}\overline{k^{(n)}_{A,ei}}k^{(n)}_{A,fj}\big)\rho^{(\phi)}_{eifj}\\
 &\leq& \frac{1}{d}\sum_{i,j}\sum_{e,f}\overline{u_{A,ei}}u_{A,fj}\rho^{(\phi)}_{eifj}\\
 &\leq& \frac{1}{d}\sum_{i,j}\rho^{(\phi)}_{iijj} = \rF_{\cE}(\rho).
\eea
Similarly, we establish an inequality $\sum_{m} p_{n,m}\rF_{\cE}(\rho_{n,m})\leq \rF_{\cE}(\rho_{n})$ for each $n$.
Therefore, it implies that $\sum_{n,m} p_{n,m}\rD^E_g(\rho_{n,m})\geq \rD^E_g(\rho)$.

\begin{center}
{\bf Appendix D: Maximum probability success in the subchannel discrimination}
\end{center}

In the subchannel discrimination,  $P_{succ}\big(\{\Psi_i\}, \{M_i\}, \rho\big)=1$ if and only if the measurement $\{M_i\}$ satisfies $\Pi_{\Psi_i(\rho)}\leq M_i$ for each $i$, where $\Pi_\sigma$ is the projection operator over all eigen-states with nonzero eigenvalues for $\sigma$.

\begin{fact}\label{fact1}
If $\sigma$ is any pure state, then  there are countless strategy $(\{\Psi_i\}, \{M_i\})$ that satisfy $P_{succ}\big(\{\Psi_i\}, \{M_i\}, \sigma\big)=1$.
\end{fact}
This is a clear fact and we can give a useful example here. Let $\{\ket{\varphi_i}\}$ is a basis of $\cH$.
Suppose that in a strategy $(\{\Psi_i\}, \{M_i\})$ operations $\{\Psi_i\}$ are defined through a series of unitary operators $U_i$ that perform $\Psi_i(\sigma) = p_iU_i\sigma U_i^\dagger = p_i\out{\varphi_i}{\varphi_i}$ with a probability distribution $(p_i)$. At this time, if measurement $M_i = \out{\varphi_i}{\varphi_i}$ is performed on the quantum states converted through $\{\Psi_i\}$, we have a success probability of $1$ regardless of the probability distribution $(p_i)_i$ in which the operations $\{\Psi_i\}$ are performed.

Then, from the definition of $\Omega_\sigma$, Eq. (9) in main text can be rewritten as
\be\label{def35}
\small{\max_{\sigma\in \cR^{\overline{max}}}\min_{\Omega_\sigma}}P_{succ}(\{\Psi_i\}, \{M_i\}, \rho).
\ee

\begin{center}
{\bf Appendix E: Proof of Theorem 7}
\end{center}

We first consider $\min_{\Omega_\sigma}P_{succ}(\{\Psi_i\}, \{M_i\},\rho)$ for the quantum state $\rho$ and for any maximum resource state $\sigma$. Here, all maximum resource states $\sigma$ are pure states and can be written as $\sigma = \out{\phi_\sigma}{\phi_\sigma}$. We already know that the measurement $\{M_i\}$ that satisfies $\Pi_{\Psi_i(\out{\phi_\sigma}{\phi_\sigma})}\leq M_i$ for any strategy $(\{\Psi_i\}, \{M_i\})$.
It implies that, for any pure state $\out{\psi}{\psi}$, the following inequality is established
\be
\tr{M_i\Psi_i(\out{\psi}{\psi})} \geq \tr{\Psi_i(\out{\psi}{\psi})}|\iinner{\phi_\sigma}{\psi}|^2
\ee
where $\ket{\psi} = \iinner{\phi_\sigma}{\psi}\ket{\phi_\sigma} + \delta \ket{\phi^\perp_\sigma}$ with $|\iinner{\phi_\sigma}{\psi}|^2+ |\delta|^2 = 1.$
Therefore, when the spectral decomposition of $\rho$ is $\rho = \sum_jq_j\out{\psi_i}{\psi_i}$, we have that
\bea
P_{succ}(\{\Psi_i\}, \{M_i\}, \rho) &=& \sum_{i,j}q_j \tr{M_i\Psi_i(\out{\psi_j}{\psi_j})}\\
 &\geq& \sum_{i,j}q_j\tr{\Psi_i(\out{\psi_j}{\psi_j})}|\iinner{\phi_\sigma}{\psi_j}|^2\\
 &=& \sum_{j}q_j|\iinner{\phi_\sigma}{\psi_j}|^2 = \rF(\sigma, \rho),
\eea
Since this inequality is established for any strategy $(\{\Psi_i\}, \{M_i\})$ , we obtain that
$$\min_{\Omega_\sigma}P_{succ}(\{\Psi_i\}, \{M_i\}, \rho)\geq \rF(\sigma, \rho).$$

Conversely, we can design a strategy for all maximum resource state $\sigma$:
Let $\{\ket{\varphi_i}\}$ is a basis of $\cH$.
Suppose that in a strategy $(\{\Psi'_i\}, \{M'_i\})$, for each $i$, operation $\Psi_i$ implemented is defined through the unitary operator $U_i$ that perform $\Psi'_i(\sigma) = p_iU_i\sigma U_i^\dagger = p_i\out{\varphi_i}{\varphi_i}$ with a probability distribution $(p_i)$, and the measurement $\{M'_i\} = \out{\varphi_i}{\varphi_i}$ is performed on the quantum states converted through $\{\Psi'_i\}$. Then, we have
\bea
&&P_{succ}(\{\Psi'_i\}, \{M'_i\}, \rho) = \sum_{j}q_j \tr{\out{\varphi_i}{\varphi_i}\Psi'_i(\out{\psi_j}{\psi_j})}\\
&&\qquad \qquad \quad = \sum_{j}q_j \tr{U_i^\dagger\out{\varphi_i}{\varphi_i}U_i U_i^\dagger\Psi'_i(\out{\psi_j}{\psi_j})U_i}\\
&&\qquad \qquad \quad = \sum_{i,j}p_iq_j|\iinner{\phi_\sigma}{\psi_j}|^2 = \rF(\sigma, \rho).
\eea
This means
$$\min_{\Omega_\sigma}P_{succ}(\{\Psi_i\}, \{M_i\}, \rho)= \rF(\sigma, \rho),$$
so, we get that
\bea
\max_{\sigma\in \cR^{\overline{max}}}\min_{\Omega_\sigma}P_{succ}(\{\Psi_i\}, \{M_i\}, \rho)&=& \max_{\sigma\in \cR^{\overline{max}}}\rF(\sigma, \rho)\\ &=&  1-\rD_g(\rho).
\eea

\end{document}